 \documentclass[aps,pre,twocolumn,groupedaddress,amsmath,amssymb]{revtex4}
\usepackage{graphicx}
\usepackage[usenames]{color}

\begin{document}


\title{A Statistical Analysis of RNA Folding Algorithms Through Thermodynamic
Parameter Perturbation}

\author{D.M. Layton${}^{1,2}$ and R. Bundschuh${}^1$\\
\small ${}^1$Department of Physics, Ohio State University, 174 W 18th Av.,
Columbus OH 43210-1106\\
\small ${}^2$Current address: University of Illinois at Urbana-Champaign,
1110 W Green St., Urbana IL 68101}

\maketitle

\newpage

\section*{Abstract}
  Computational RNA secondary structure prediction is rather well established.
  However, such prediction algorithms always depend on a large number of
  experimentally measured parameters. Here, we study how sensitive structure
  prediction algorithms are to changes in these parameters. We find that
  already for changes corresponding to the actual experimental error to which
  these parameters have been determined 30\% of the structure are falsly
  predicted and the ground state structure is preserved under parameter
  perturbation in only 5\% of all cases. We establish that base pairing
  probabilities calculated in a thermal ensemble are a viable though not
  perfect measure for the reliability of the prediction of individual
  structure elements.  A new measure of stability using parameter perturbation
  is proposed, and its limitations discussed.

\section*{Introduction}

In an endeavor to understand the workings of any organism, one cannot ignore
the importance of ribonucleic acids (RNA)~\cite{science_paper}. RNA molecules
transmit genetic information through the cell. They are also intimately
involved in many important biological processes such as translation,
regulation, and splicing. In addition to its importance for today's organisms
RNA is also an interesting molecule to study due to its likely role as a major
player during the origin of life~\cite{rnaworld}.

The function of a given RNA is determined by its physical structure. This
structure is encoded in the sequence of the four nucleotides (or bases) A, U,
G, and C from which each RNA molecule is composed. Determining that structure
in the laboratory is a laborious, and often unsuccessful, undertaking. Thus,
it has become an interdisciplinary task to determine these structures from the
sequences alone.

The encoding of a structure in the sequence is realized by specific
interactions between the bases. The by far most important of these
interactions is the formation of A--U and G--C base pairs, also known as
Watson-Crick pairs. With the formation of each base pair, the Gibbs free
energy of the structure is lowered, and thus, the structure's stability is
increased. Since the sequence of bases that defines the RNA is finite, the
number of possible structures into which a given RNA can fold is also finite.
The most thermodynamically likely structure to be formed
is the structure
with the lowest free energy known as the minimum free energy structure.

Although the number of possible structures for a given
sequence is enormous, computer algorithms, such as the Vienna
Package~\cite{hofac} or MFOLD~\cite{zuker}, can find the minimum free
energy structure or the full partition function of the ensemble of all
structures given a sequence in a time that is proportional to the
third power of the sequence length due to a recursive
relationship~\cite{dege68,wate78,mccaskill}. The
problem with these algorithms lies in the calculation of the free energies of
the structures. The contribution to the free energy attributed to a base
pairing or the formation of various substructures such as various kinds of
loops is measured experimentally and used as parameters in the RNA folding
algorithm.  For various reasons, these parameters contain errors.
On the one hand, several effects such as steric interactions
between different regions of the structure, pseudo-knots, base triplets,
or even interactions with proteins or other RNA molecules are not reflected
at all in the underlying free energy model. All these effects result
in systematic errors in the free energy parameters. On the other hand,
there are ordinary, non-systematic
errors of measurement associated with these parameters as well.
Thus, while the algorithm guarantees to find the minimum energy structure
within the energy model provided by the experimentally determined parameters,
this structure does not have to be the true (and thus biologically realized)
minimum free energy structure.

The goal of this paper is not to discuss the causes of these experimental
errors but simply, to investigate the consequences these errors have regarding
structure prediction. Our approach is to randomly modify the measured free energy
parameters within a range comparable to the experimental errors and to record
how much the predicted minimum free energy structures change. We find that
already around 30\% of the structures are changed if the free energy
parameters are varied within the experimental error. While this is a rather
sobering result, we at least find that base pairing probabilities evaluated in
a thermal ensemble are very good priors to estimate which parts of the
structure prediction are reliable and which are not.


\section*{Materials and Methods}\label{sec_secondary}

In this section we will discuss how RNA structure prediction algorithms work
and how we model experimental error in the free energy parameters. This will
provide the necessary background for our study.

\paragraph{RNA secondary structure:}
The strongest interaction between the bases of an RNA molecule is the
formation of Watson-Crick base pairs. Therefore, one distinguishes two
levels of structure, namely secondary and tertiary structure (with the
primary structure just being another name for the sequence of the
molecule.) A secondary structure is defined as the collection of all
base pairs that have formed without regard to any spatial organization
of the backbone. The tertiary structure then includes the actual
spatial organization and elements formed by less stable interactions
than base pairing such as base triplets, backbone contacts mediated by
divalent ions, etc. Since base pairing is energetically more important
than the other interactions, it is meaningful to talk about the
secondary structure of an RNA molecule without considering the
tertiary structure~\cite{tinocobustamante}. This is the point of view
of the algorithms that we will study here; therefore we also will
discuss secondary structure only for the remainder of this paper.

In order to make secondary structure prediction computationally feasible, it
is necessary to exclude so-called pseudoknots from the allowed secondary
structures. Such a pseudoknot exists if bases $i$ and $j$ form a base pair and
bases $k$ and $l$ form a base pair and these two base pairs are nested as
$i<k<j<l$ or $k<i<l<j$. While these pseudoknots do appear in biological
structures, they are bound to be short due to kinetic constraints. Thus,
they can be omitted in secondary structure
prediction~\cite{tinocobustamante} and be considered part of the tertiary
structure of a molecule.

\begin{figure}[h]
\includegraphics[width=84mm]{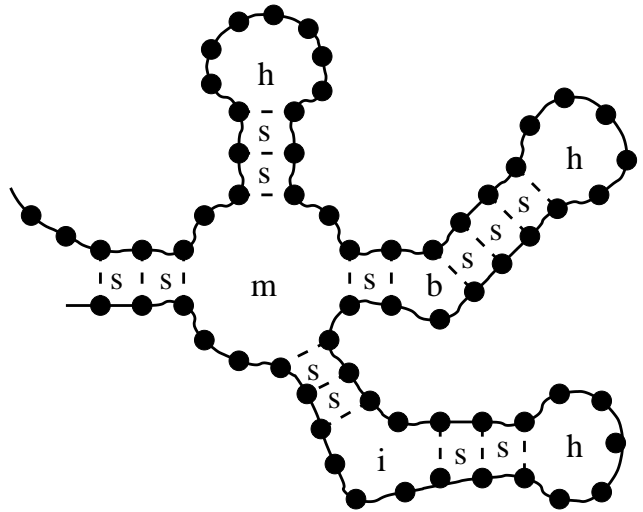}
\caption{Schematic of an RNA secondary structure. The solid line
represents the backbone of the molecule while the dashed lines
symbolize base pairs. Each such structure can be decomposed into
stacking loops (s), bulges (b), interior loops (i), hairpin loops
(h), and multi-loops (m) as indicated.} \label{rna}
\end{figure}

\paragraph{Energy model:}
A secondary structure as defined above can be drawn as shown in Fig.~\ref{rna}.
It can be decomposed into a large number of loops that are either
stacking loops, bulges, interior loops, hairpins, or multiloops as also
indicated in the figure. The main assumption of the generally accepted free
energy model is that the total free energy of a secondary structure is the sum
of independent contributions from all of its loops. These loop contributions
depend on the identity of the bases in a loop and on the length of the
loop. E.g., the free energy of a stacking loop depends on the identity of all
four bases that form the two base pairs a stacking loop is formed by leading
to (after taking into account symmetry) $21$ different free energy parameters
for stacking loops (if in addition to G--C, and A--U also the wobble base pair
G--U is allowed). For short bulges and interior loops the number of parameters
increases by a factor of four for every unpaired base in the loop. Longer
loops are typically only characterized by their length and by the identity of
the unpaired bases immediately next to the base pairs defining the loop in
order to avoid an explosion of parameters. Nevertheless, a complete free
energy model is described by on the order of a thousand parameters that are
determined experimentally~\cite{mathews}. Since all these parameters are true
free energies, i.e., differences of energetic contributions such as chemical
binding energy and bending energy and entropic contributions from the
integrated out spatial degrees of freedom of the backbone and the surrounding
water, each parameter depends on the temperature. In our study we will keep the
temperature constant at physiological $37^o$C.

\paragraph{Perturbations of the energy model:}
In order to study the sensitivity of structure prediction to thermodynamic
parameters, one must perturb the parameters. For simplicity we assume that the
error in the parameters is roughly Gaussian distributed. We model these errors
by adding to every single free energy parameter a Gaussian random variable
with mean zero and standard deviation, $\epsilon$, i.e., with a probability
density
\begin{equation}
\rho(x)=\frac{1}{\sqrt[]{2\pi}\epsilon}e^{-\frac{x^2}{2\epsilon^2}}.
\end{equation}
In doing so we take great care to preserve the inherent symmetry
in the parameters (e.g. the energy of a stacking loop made from a
GC-pair and an AU-pair (--GA-- paired with --UC--) being equal to
the energy of a stacking loop made from an UA-pair and a CG-pair,
i.e., --UC-- paired with --GA--). The parameter $\epsilon$ serves
as a measure of the magnitude of the experimental error inherent
in the parameters.

We will explore a whole range of different values for the
parameter $\epsilon$ to understand the sensitivity of predicted
structures to perturbations of the free energy parameters. In order to
get an idea what experimental errors on the free energy parameters are
in reality, it is most illustrative to look at the stacking energies
since stacking energies have been measured in many different
laboratories.  In the case of the more studied DNA stacking energies
seven independent measurements have been systematically
compared~\cite{santalucia}.  This study reports in addition the
detailed free energy parameters the stacking energies averaged over
the different types of stacking for each of the seven independent
experiments. If we take the average of these averages we find it to be
$-1.4\pm0.3$kcal/mol. Since the experimental procedure for the
determination of RNA free energy parameters is the same as for DNA
free energy parameters we conclude that a good estimate for the
experimental error is $0.3$kcal/mol.  Another indication that this is
the order of magnitude of the experimental error of the stacking free
energies is that the additive free energy model itself is
experimentally known to break down at this level of
precision~\cite{kierzek}. This implies that this uncertainty is not
due to a lack of experimental techniques of higher precision (which
could in principle be overcome by new experimental developments) but
that these uncertainties are unavoidable on principle grounds. Since
we do not have very good estimates of the experimental error of the
other free energy parameters we uniformly apply the same error
estimate to all free energy parameters.

\newpage
\begin{table*}[tb]
\begin{tabular}{l|r|r|r|r|r|r|r|r|r|r|r|r|r}   \hline\hline
Accession& Length
         & \multicolumn{2}{c|}{ $\epsilon=0.02\frac{\mathrm{kcal}}{\mathrm{mol}}$}
         & \multicolumn{2}{c|}{ $\epsilon=0.12\frac{\mathrm{kcal}}{\mathrm{mol}}$}
         & \multicolumn{2}{c|}{ $\epsilon=0.22\frac{\mathrm{kcal}}{\mathrm{mol}}$}
         & \multicolumn{2}{c|}{ $\epsilon=0.32\frac{\mathrm{kcal}}{\mathrm{mol}}$}
         & \multicolumn{2}{c|}{ $\epsilon=0.42\frac{\mathrm{kcal}}{\mathrm{mol}}$}
         & \multicolumn{2}{c}{ $\epsilon=0.52\frac{\mathrm{kcal}}{\mathrm{mol}}$} \\ \hline
Random & 190 & 0 & 100.0\% & 147& 10.5\%& 314 & 7.4\%& 576 & 7.3\% & 829 & 1.1\% & 983 & 0.5\% \\
Random & 210 & 1 & 95.7\% & 291& 2.3\%& 541 & 2.2\%& 846 & 1.0\% & 977 & 0.6\% & 989 & 0.1\% \\
AJ228695 & 227 & 0 & 100.0\% & 20& 46.8\%& 128 & 15.8\%& 350 & 5.9\% & 619 & 2.1\% & 829 & 1.1\% \\
Random & 230 & 15 & 34.0\% & 195& 5.5\%& 445 & 3.5\%& 855 & 1.6\% & 994 & 0.1\% & 991 & 0.2\% \\
AJ228705 & 248 & 0 & 100.0\% & 38& 58.1\%& 223 & 18.4\%& 574 & 6.4\% & 873 & 1.8\% & 965 & 0.6\% \\
U83261   & 243 & 4 & 98.0\%  & 49& 23.1\%& 241 & 8.8\% & 591 & 3.4\% & 880 & 1.0\% & 976 & 0.2\% \\
Y13474   & 256 & 0 & 100.0\% & 23& 86.1\%& 167 & 32.7\%& 491 & 6.0\% & 794 & 1.2\% & 940 & 0.3\% \\
Random & 270 & 4 & 76.2\% & 353& 3.4\%& 681 & 1.3\%& 926 & 0.2\% & 993 & 0.2\% & 999 & 0.1\% \\
Random & 290 & 1 & 99.9\% & 134& 23.1\%& 513 & 1.7\%& 908 & 0.2\% & 996 & 0.1\% & 995 & 0.1\% \\
Random & 330 & 9 & 58.8\% & 327& 7.1\%& 909 & 1.1\%& 986 & 0.2\% & 997 & 0.1\% & 999 & 0.1\% \\
Random & 350 & 2 & 79.5\% & 243& 8.1\%& 581 & 2.2\%& 990 & 0.1\% & 998 & 0.1\% & 999 & 0.1\% \\
M38691 & 376 & 6 & 61.5\% & 516 & 2.6\% & 941 & 0.2\% & 996 & 0.1\% & 1000 & 0.1\% & 1000 & 0.1\% \\
Random & 390 & 6 & 27.1\% & 321& 11.8\%& 846 & 0.5\%& 997 & 0.1\% & 999 & 0.1\% & 1000 & 0.1\% \\
V01416   & 426 & 7 & 40.4\% & 219 & 9.4\% & 749 &  1.7\% & 977 & 0.4\% & 999 & 0.2\% & 1000 & 0.1\% \\
 \hline \hline
\end{tabular}
\caption{Frequency at which the ground state is predicted (right)
and the number of alternate structures predicted (left) as a
function of the parameter perturbation $\epsilon$. Entries labeled
``Random'' are randomly generated squences. All others are group I introns.}
\label{grdstprob}
\end{table*}

\section*{Results}\label{sec_structurediff}

If the predicted minimum free energy (mfe) structure, also referred to as the
ground state, has a far lower free energy than any alternative structure, the
ground state is said to be thermodynamically stable. For the purposes of this
paper, we are interested in another type of stability. We will call a
structure unstable with respect to parameter perturbation should the predicted
mfe structure require a strict adherence to one or more thermodynamic
parameters in order to remain the predicted mfe structure. We will quantify
this stability in the following in two different ways.

\paragraph{Distance of structures:}
The first way we study this instability is by looking at what fraction of a
structure is still predicted correctly once parameters are perturbed.  To this
end, we need some quantitative method of comparison for structures.  In this
study, we will use the normalized tree distance~\cite{hofac} to quantify the
amount by which mfe structures at different free energy parameter choices differ. We
convinced ourselves that other solely structure based measures such as the
string distance~\cite{hofac} lead to the similar results as the ones
presented here. The tree distance is based on a metric that views a secondary
structure as being defined by a tree diagram where the leafs of the tree are
the bases and the topology of the tree represents the structure in an
intuitive way. The tree distance is then defined as the number of elementary
operations on the tree such as cutting a branch and attaching it at a
different place in the tree.  For a more detailed description of this
difference measure the reader is referred to Ref.~\cite{fontana}. Since the
tree distance is a number between zero (for identical structures) and the
length of the sequence, we rescale the tree distance by the length of the
sequence. This scaling allows us to compare the stability of sequences of
different lengths and permits a more intuitive interpretation of the data.
For example, a scaled distance of $0.2$ stands for a 20\% difference in
structure.

For our study we choose a series of natural
sequences (namely group I introns) with length
varying between 227 and 685, that is, RNA sequences which have been
observed in biological systems. For each of these sequences the mfe
structure is predicted by the computer algorithm from the Vienna
package~\cite{hofac} using the accepted experimentally measured
parameters~\cite{mathews}.  In addition, we determine the mfe
structures of the same sequences for a hundred sets of randomly
perturbed energy parameters for each uncertainty $\epsilon$.  We
calculate the scaled tree distance between each mfe structure
calculated with perturbed parameters and the corresponding mfe
structure obtained with unperturbed parameters (i.e. the
experimentally measured values without alteration) and average these
distances over the $100$ realizations for each sequence and each~$\epsilon$.

\begin{figure}[h]
\includegraphics[width=84mm]{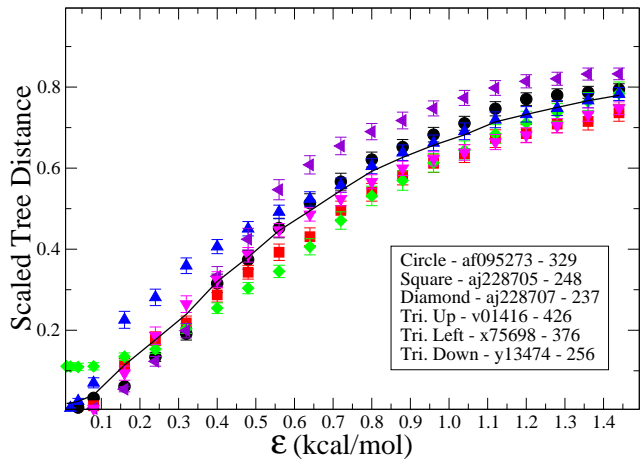}
\caption{The average scaled tree distances of various natural
sequences
  for different perturbations of the parameters $\epsilon$. The error bars
  denote the statistical error after averaging over $100$ different
  realizations of perturbed free energy parameter sets. The solid
  line shows the average over all sequences. It becomes obvious that already
  at the experimental uncertainty of \protect$\epsilon\approx0.3$kcal/mol
  about 30\% of the structure is predicted unreliably.}
\label{natdist}
\end{figure}

Fig.~\ref{natdist} shows these averaged distances as a function of the
perturbation parameter $\epsilon$. One should note the overall instability of
the structure prediction of these natural sequences.
As can be seen in Fig.~\ref{natdist}, there is already
a significant deviation of about 30\% from the ground state structure at
$\epsilon\approx0.3$kcal/mol which roughly corresponds to the actual
experimental error of the parameters~\cite{santalucia}. For $\epsilon\approx
1$kcal/mol already half of the structure can no longer be predicted.

\paragraph{Ground state probability:}

An alternative, and as it turns out even more sensitive, measure of stability
is the probability that the ground state structure will be the predicted
structure at a given perturbation $\epsilon$.  To estimate this quantity we
determine the mfe structure of our sequences for $1000$ sets
of randomly perturbed parameters for each perturbation strength $\epsilon$ and
classify the parameter sets according to the mfe structures. Since we catalog
the parameter sets which produce each structure, the frequency at which the
``correct'' (i.e., calculated with the accepted free energy parameters)
structure occurs can be determined, as well as the number of alternative
structures possible at a given~$\epsilon$.

As can be seen in table 1, the ground state structure is most
likely not the true structure for one sequence for $\epsilon$ as
small as $0.02$, and for all sequences for $\epsilon$ of $0.22$.
At the experimental error rate, i.e., $\epsilon\approx0.32$, only
a mere 5\% of the parameter perturbations reproduce the same
ground state.
For this part of the
study we also include data on randomly generated sequences of varying length
in addition to data on the group I introns used in the rest of this
manuscript. Table 1 shows that the ground state stability for these sequences
is even worse than for the group I introns.
 From this data, one can see that the current error
in the thermodynamic parameters casts serious doubt upon the
structural predictions made by folding algorithms. Since
this error is at least to a good part due to fundamental limitations in
the energy model and thus cannot be significantly reduced just by better
measurements, predictions
from folding algorithms should never be taken at face value
but always be subjected to critical crosschecking.

\paragraph{Reliability estimation:}
\begin{figure*}[tb]
\includegraphics[width=178mm]{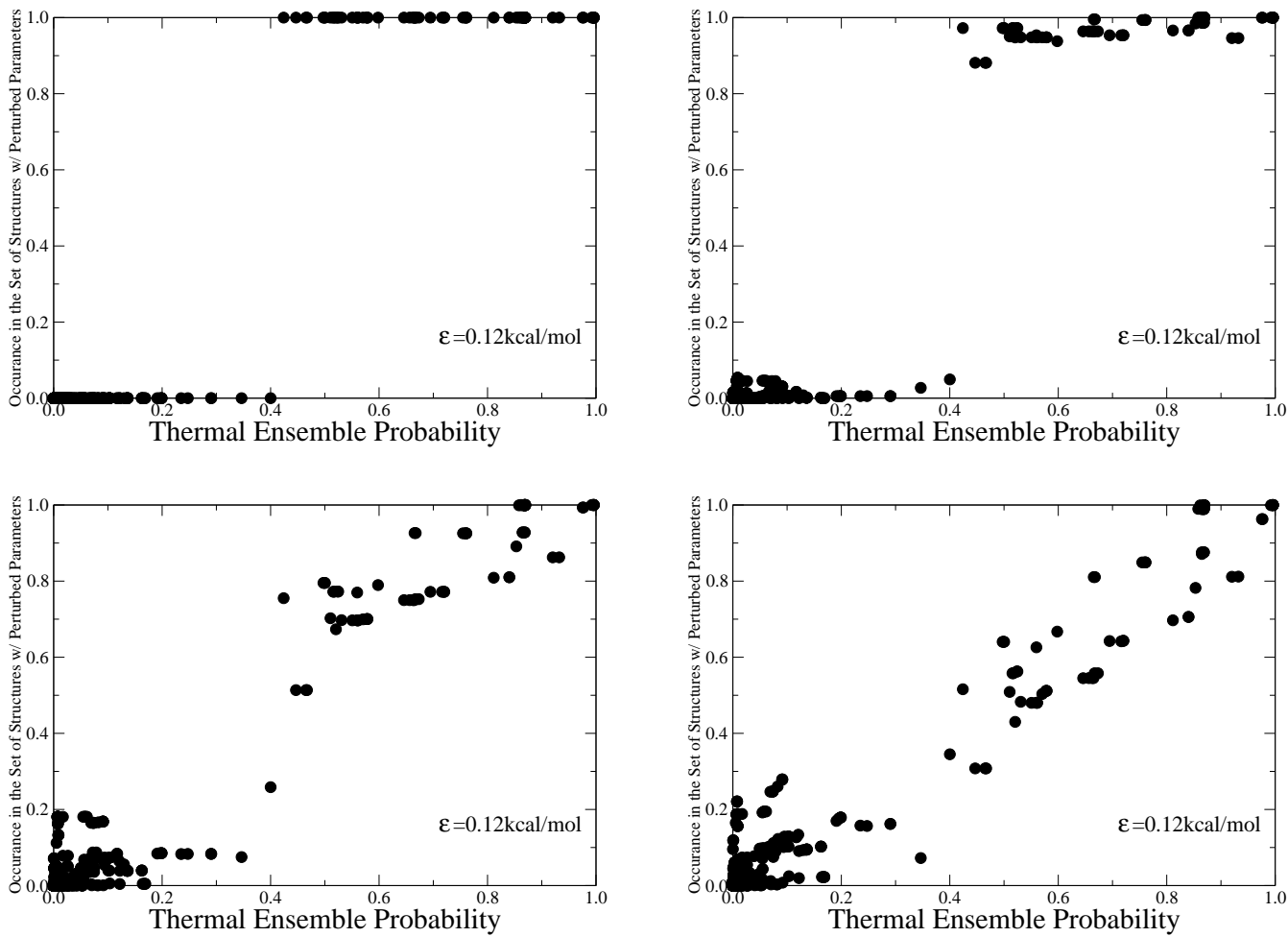}
\caption{Correlation between the probability to find a base pair
in an mfe structure with perturbed parameters and the probability
of the base pair in the thermal ensemble for the sequence with
accession number Y13474.  Each point represents a base pair and
its position represents its respective probabilities of forming.
The different plots show how the two probabilities become more
correlated as the strength \protect$\epsilon$ of the parameter
perturbations is increased.} \label{eprog}
\end{figure*}

Given that we obviously have to live with the fact that on the order of 30\%
of a secondary structure will be incorrectly predicted just because of the
uncertainties in the free energy parameters, the question comes up
if it is at least possible to find out which parts of the structure are the
reliable ones and which are the unreliable ones. If the RNA secondary
structure prediction algorithm is used to calculate the full partition
function for a given sequence instead of the mfe structure only, it can
assign to every pair $(i,j)$ of bases a probability $p_{i,j}$ that these two
bases are paired within a thermal ensemble~\cite{mccaskill}. Since these
probabilities can be calculated in the same time as the mfe structure, these
probabilities are a convenient measure for the reliability of the prediction
of an individual base pair. However, it is not a priori clear if a high
probability in the thermal ensemble corresponds to stability with respect to
uncertainties in the free energy parameters.

To study if the thermal probabilities have any meaning for the stability of a
base pair with respect to parameter changes, we compare the thermal ensemble
directly to an ensemble of mfe structures calculated with perturbed
parameters.  To this end we calculate the mfe structure for a given sequence
for $1000$ different sets of perturbed free energy parameters. We catalog the
resulting base pairs, and determine the frequency with which they occur. Then,
we compare this frequency to the thermal ensemble probability $p_{i,j}$
calculated at the accepted parameters for each individual base pair.

The comparison of the base pair frequencies versus the thermal ensemble
probabilities is shown in Fig.~\ref{eprog} for a representative sequence (we
convinced ourselves that the results are qualitatively similar for all
sequences). We observe that at small $\epsilon$, since few or no alternative
structures are predicted, the plot appears to be very much a step function;
base pairs which the thermal ensemble predicts to have a significant
probability ( $\approx$ 40\% or more) occur while less likely base pairs do
not. As $\epsilon$ is increased, more alternative structures begin to appear,
and one can see the edges of the step begin to smooth. By
$\epsilon=0.32$kcal/mol a strong correlation is apparent even though there is
a clear spread. If increased beyond
$\epsilon=0.32$kcal/mol, the correlation
between the base pair frequency and the thermal ensemble probability differs
in no significant qualitative way. From these correlations we
deduce that as the level of parameter perturbations
reaches the value of $\epsilon=0.32$kcal/mol the pool of alternative
secondary structures minimizing perturbed energy parameters and the
pool of suboptimal structures probed in the thermal ensemble become
similar. This is in a way surprising since the thermal energy itself
is about $0.6$kcal/mol and thus much smaller than the perturbation
of the total free energy of a structure obtained by perturbing every
free energy parameter by $0.32$kcal/mol. It might imply that the
base pair probability of an individual base pair is only sensitive to
perturbations of a few key free energy parameters that delineate
different low free energy structures from each other. Whatever the reason for
the observed
correlation, we can conclude that the easily calculable thermal probabilities
are a good estimate for the sensitivity of a
given base pair to parameter perturbations.

\section*{Discussion}

We studied the sensitivity of RNA secondary structure prediction
to perturbations of the free energy parameters. The main result is
that if the free energy parameters are perturbed within a range
that is supposed to be the experimental uncertainty with which
these parameters have been determined, about 30\% of the structure
turns out to be unreliable and the chance of predicting the same
ground state as with the unperturbed parameters is only 5\% even
for moderately sized sequences with lengths up to 426. Given this
imprecision, we found that at least base pairing probabilities
calculated in a thermal ensemble are reasonably well correlated
with the probabilities that a base pair will be unaffected by
uncertainties in the free energy parameters. 
These results support the commonly employed method of using
thermal ensemble    
probabilities to sort out which parts of the structure can be
trusted and which cannot. 
However, although calculation of the thermal
ensemble~\cite{zuker1998,mathews2004} 
is expedient, it offers only knowledge of how base pairings will 
behave on a individual basis, and not how they will behave in concert. 
 The ground state probability method gives
one not only a probabilistic measure of the accuracy of the
prediction, but also all the probable alternative structures and
some gauge of their likelihood of being the {\em true} structure.
Should one have the computer power, one should always check the ground
state and individual base pair probabilities using the method
outlined in this paper. Doing so is imperative if the thermal ensemble 
suggests a dubious structure prediction.
The amount of time sacrificed for the additional information is
dependent upon how accurate one wishes the probabilities to be.
With advances in computer chip technology, the extra factor of
$100$ to $1000$ in computation time involved in using the ground
state probability as opposed to using the thermal ensemble may
soon become a more practical investment in cases where it is of
big importance to know in addition to the predicted structure
which parts of the structure are likely to be correctly predicted
and which should be discarded as simple artifacts of the
imprecisions of the free energy model.

\section*{Acknowledgments}

This research was supported by the Research Experience for
Undergraduates programs of the National Science Foundation through
grant number PHY-0242665.


\end{document}